\documentclass[a4paper]{appolb}
\usepackage{graphicx}
\usepackage{amsmath}
\usepackage{multirow}
\usepackage{cite}

\begin{document}
\title{$X(3872)$ production in high energy heavy ion collisions}
\author{A.~Mart\'inez~Torres, K. P. Khemchandani, F. S. Navarra, M. Nielsen
\address{Instituto de F\'isica, Universidade de S\~ao Paulo, C.P. 66318, 05389-970 S\~ao 
Paulo, SP, Brazil.}
\\
Luciano M. Abreu
\address{Instituto de F\'isica, Universidade Federal da Bahia, 40210-340, Salvador, BA, Brazil.}}
\maketitle
\begin{abstract}
We show the results obtained for the cross sections of the processes $\bar D D$, $\bar D^* D$, $\bar D^* D^*\to\pi X(3872)$, information which is necessary to determine the $X(3872)$ abundance in heavy ion collisions. Our formalism is based on the generation of $X(3872)$ from the interaction of the hadrons $\bar D^0 D^{*0} - \textrm{c.c}$, $D^- D^{*+} - \textrm{c.c}$ and $D^-_s D^{*+}_s - \textrm{c.c}$ an the calculation of the $X\bar D^* D^*$ anomalous vertex considering $X(3872)$ as a molecule of the above hadrons channels.  
\end{abstract}

\section{Introduction}
The $X(3872)$ (from now onwards simply $X$) was reported a decade ago by the Belle collaboration in the reaction $B^{\pm}\to K^{\pm}\pi^+\pi^- J/\psi$~\cite{Choi}. After this finding,  different collaborations  
\cite{Acosta,Abazov,Aubert} confirmed this state and very recently the spin-parity quantum numbers of $X$ have been confirmed to be $1^{++}$~\cite{Aaij}. 

During this time, several theoretical models have been proposed  to describe the properties of this state, considering it as a charmonium state, a tetraquark, a $D - \bar{D}^*$ hadron molecule  
and a mixture between a charmonium and a molecular component~\cite{Tornqvist,Close,Swanson,Braaten,Daniel2,Matheus, Dong,Daniel4,Coito,roma14}. In spite of the effort of these numerous groups, the properties of this particle are not yet well understood and represent a challenge both for theorists and experimentalists.  

In a different frontier of physics, collaborations like RHIC and LHC have devoted a significant effort to study the Quark Gluon Plasma (QGP). It is now a well accepted fact that in high energy heavy ion collisions a deconfined medium is created:  the quark gluon plasma (QGP) \cite{Arsene,Adams}. The formation of the quark gluon plasma phase increases the number of produced $X$'s and the predicted abundance is very sensitive to the structure of $X$ (tetraquark, molecular, etc.)~\cite{Cho1,Cho2,Cho3}.  In this way, heavy ion collisions can be used to obtain information about exotic charmonium states as $X$. 

However, due to the rich hadronic environment  present in the plasma, the $X$ can be destroyed/produced in collisions with ordinary hadrons, fact which can affect to its abundance. In Ref.~\cite{Cho3}, the hadronic absorption cross section of the $X$ by mesons like $\pi$ and $\rho$ was evaluated  for the processes $\pi X \to D\bar D$, $D^*\bar D^*$ and $\rho X\to D\bar D$,  $D\bar D^*$, $D^*\bar D^*$. Using these  cross sections, the variation of the $X$ meson abundance during the expansion of the hadronic matter was computed with the help of a kinetic equation with gain and loss terms. The results turned out to be  strongly dependent on the quantum numbers of the $X$ and on its structure.

The present work is devoted to introduce two improvements in the calculation of cross sections performed in Ref.~\cite{Cho3}.  
The first and most important one is the inclusion of the anomalous vertices $\pi D^* D^*$ and $X \bar D^* D^*$, which were neglected 
before. With these vertices new reaction channels become possible, such as  $\pi X\to D\bar D^*$, and the inverse process
$D\bar D^* \to \pi X$. The second  improvement is the inclusion of the charged components of the $D$ and $D^*$ mesons which couple to the $X$~\cite{Daniel4}. 

\section{Formalism}

\begin{figure}[h!]
\centering
\includegraphics[width=0.7\textwidth]{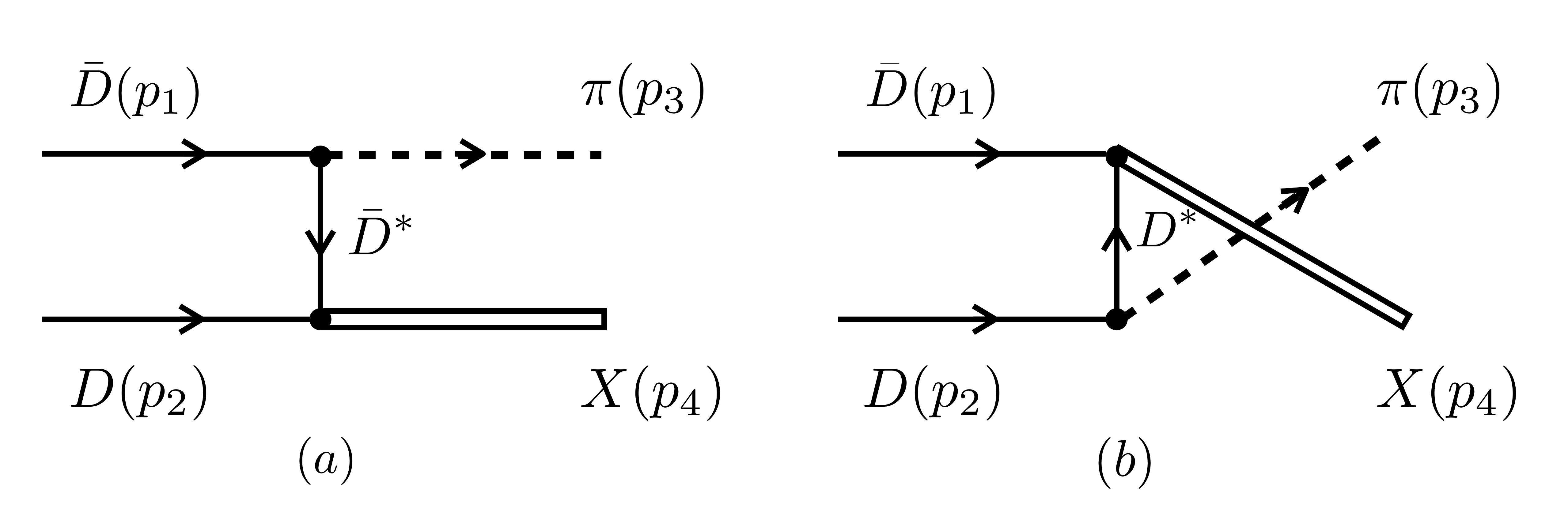}
\caption{Diagrams contributing to the process $\bar D D\to \pi X$.}\label{DbarD}
\end{figure}

\begin{figure}[h!]
\centering
\includegraphics[width=0.65\textwidth]{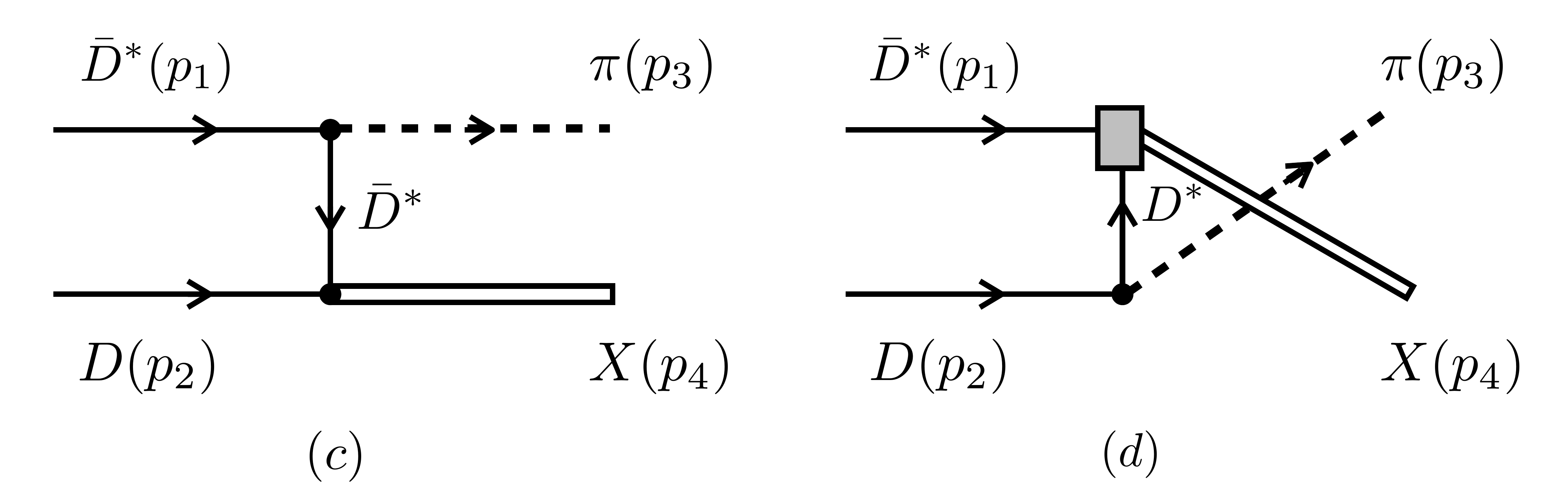}
\caption{Diagrams contributing to the process $\bar D^* D\to \pi X$. The diagram containing a filled box is calculated by summing the set of diagrams shown in Fig.~\ref{blob}, as explained in the text.}\label{DbarstarD}
\end{figure}

\begin{figure}[h!]
\centering
\includegraphics[width=0.65\textwidth]{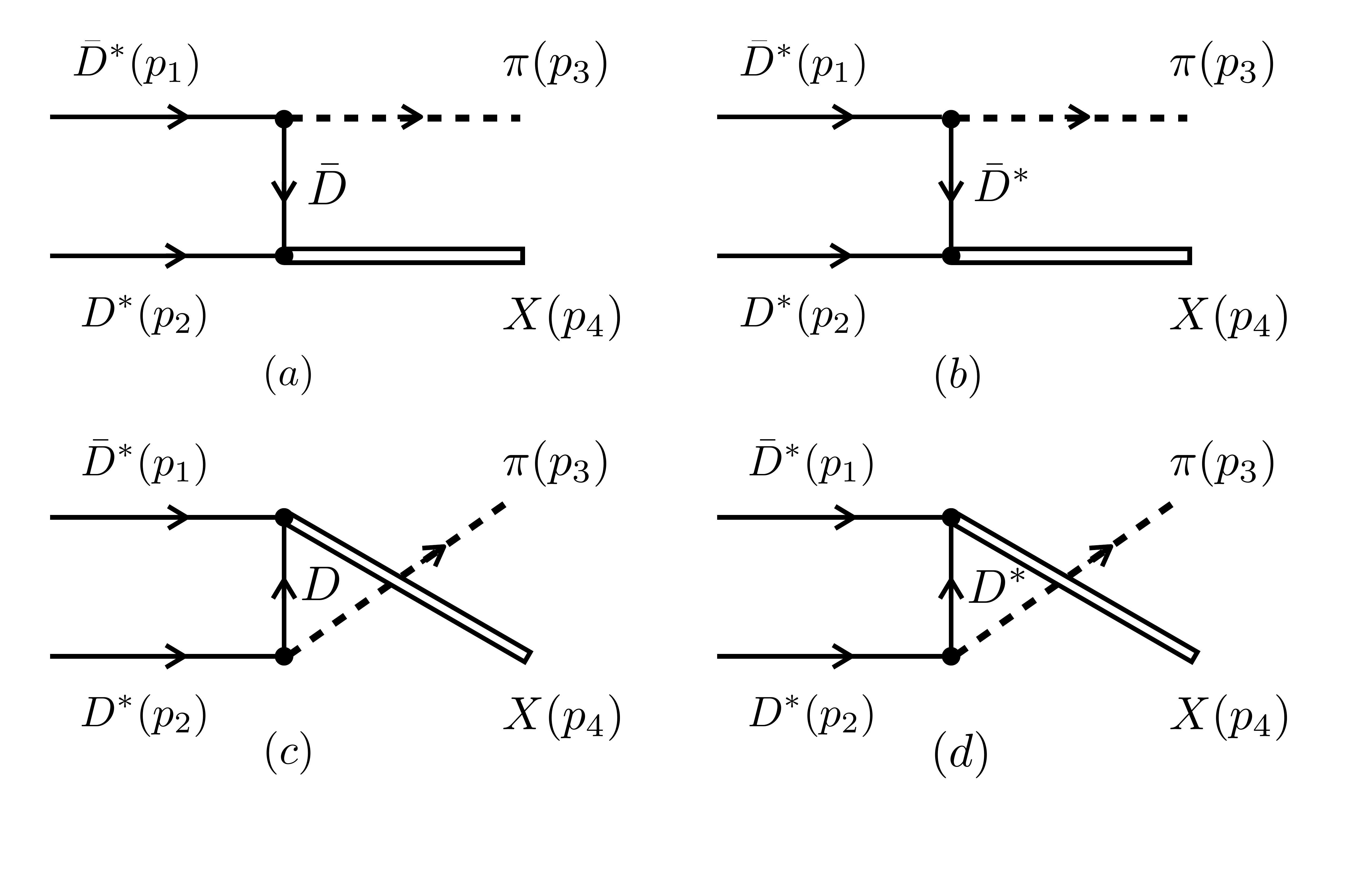}
\caption{Different diagrams contributing to the reaction $\bar D^* D^*\to\pi X$.}\label{DstarDstar}
\end{figure}

We consider the diagrams shown in Figures~\ref{DbarD},~\ref{DbarstarD} and \ref{DstarDstar} to calculate the cross sections for the processes $\bar D D$,  $\bar D^* D$, $\bar D^* D^*\to \pi X$,
respectively.

\begin{figure}[h!]
\centering
\includegraphics[width=0.9\textwidth]{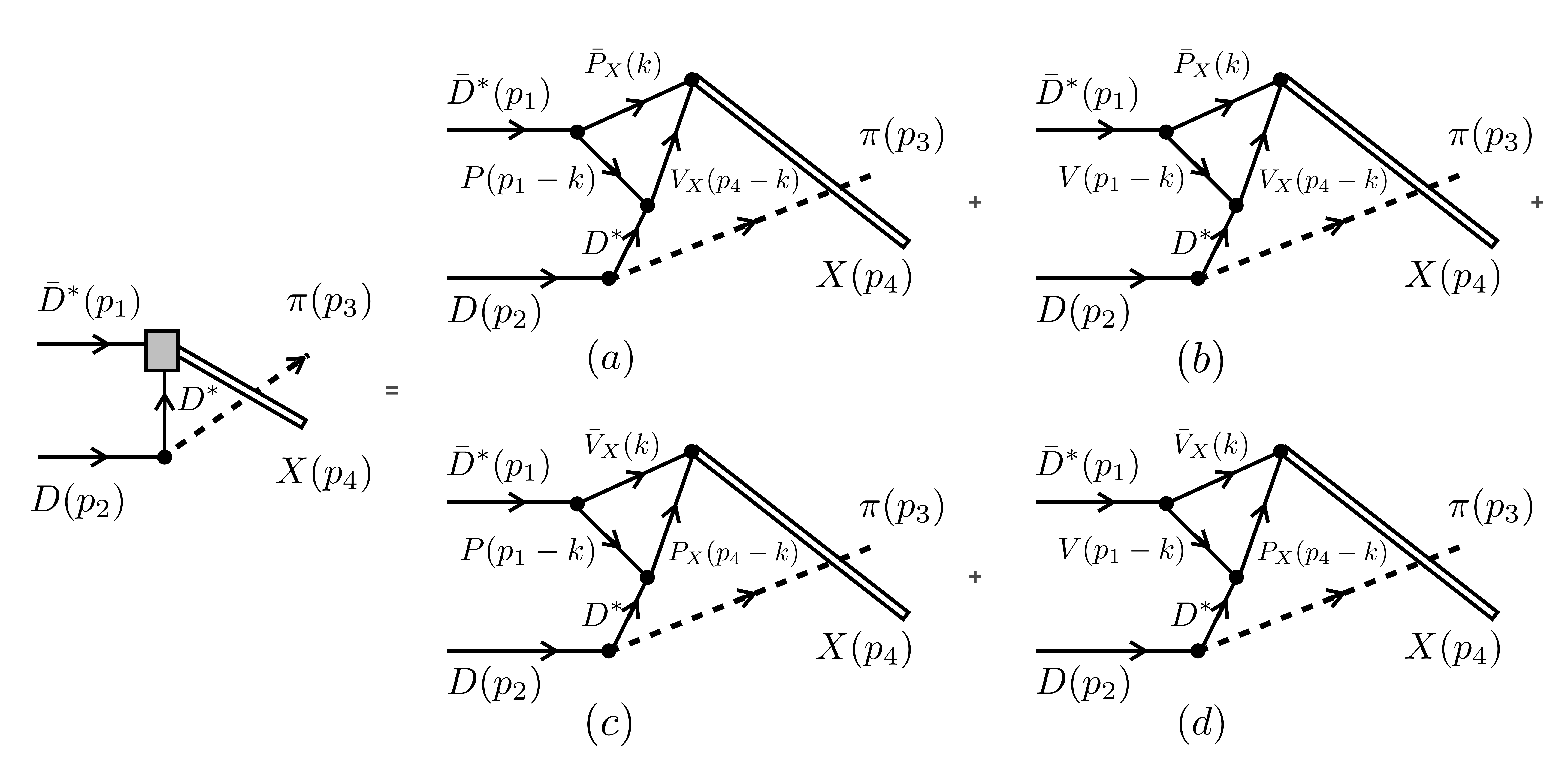}
\caption{Diagrams considered for the determination of the filled box shown in Fig.~\ref{DbarstarD}. The hadrons $P_X$ and $V_X$ represent the pseudoscalars and vectors coupling to the state $X$, while $P$ and $V$ are any
pseudoscalar and vector meson which can be exchanged conserving different quantum numbers. For a list of the different exchanged hadrons considered here see Ref.~\cite{our}.}\label{blob}
\end{figure}

To determine the contribution from the box in Fig.~\ref{DbarstarD} we evaluate the diagrams in Fig.~\ref{blob} following the model of Refs.~\cite{Daniel2,Daniel4,Francesca2} in which $X$ is generated from the interaction of $\bar D^0 D^{*0}-\textrm{c.c}$, $D^- D^{*+}-\textrm{c.c}$ and $D^-_s D^{*+}_s-\textrm{c.c}$.

The isospin-spin averaged production cross section for the processes $\bar D D, \bar D^* D, \bar D^* D^*\to \pi X$, in the center of mas (CM) frame can be determined as
\begin{align}
\sigma_r(s)=\frac{1}{16\pi\lambda(s,m^2_{1i,r},m^2_{2i,r})}\int^{t_{\textrm{max,r}}}_{t_\textrm{min,r}}dt\overline{\sum\limits_{\textrm{Isos},\textrm{spin}}}\left |\mathcal{M}_r(s,t)\right|^2,\label{cross}
\end{align}
where $r=1,2,3$ is an index indicating the reaction considered, $\sqrt{s}$ is the CM energy, and $m_{1i,r}$ and $m_{2i,r}$ represent the masses of the two
particles present in the initial state $i$ of the reaction $r$. The function $\lambda(a,b,c)$ in Eq.~(\ref{cross}) is the K\"allen function, $t_\textrm{min,r}$ and $t_{\textrm{max,r}}$ correspond to the minimum and maximum values, respectively, of the Mandelstam variable $t$ and $\mathcal{M}_r$ is the reduced matrix element for the process $r$.  The symbol $\overline{\sum\limits_{\textrm{spin},\textrm{Isos}}}$ represents the sum over the isospins and spins of the particles in the initial and final state, weighted by the isospin and spin degeneracy factors of the two particles forming the initial state for the reaction $r$.

For the mathematical expressions of the amplitudes related to the diagrams shown in Figs.~\ref{DbarD},~\ref{DbarstarD} and \ref{DstarDstar} we refer the reader to Ref.~\cite{our}. Here we just 
discuss the results found using these amplitudes for the different cross sections.

\section{Results}
\begin{figure}
\centering
\includegraphics[width=0.45\textwidth]{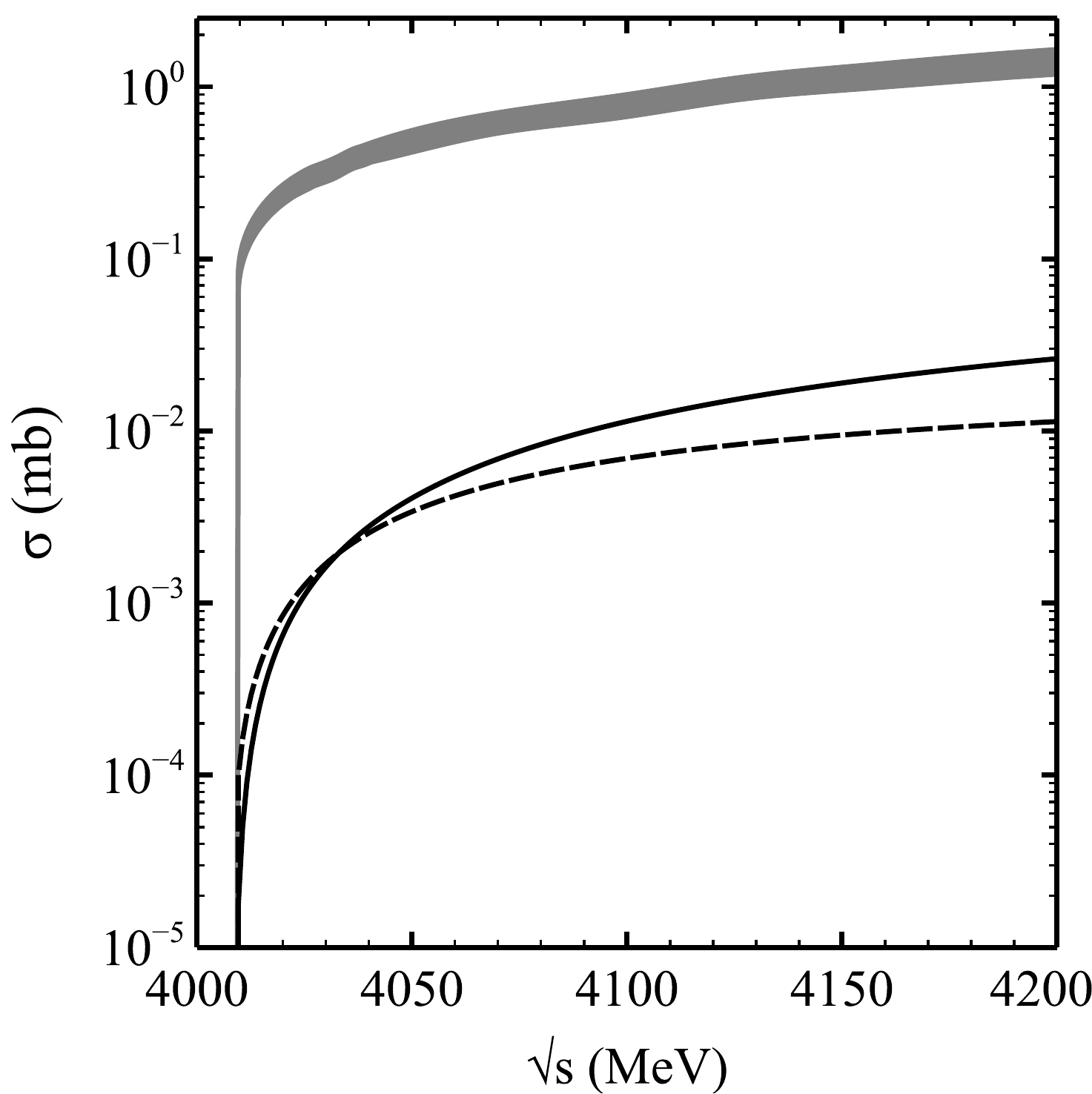}
\includegraphics[width=0.45\textwidth]{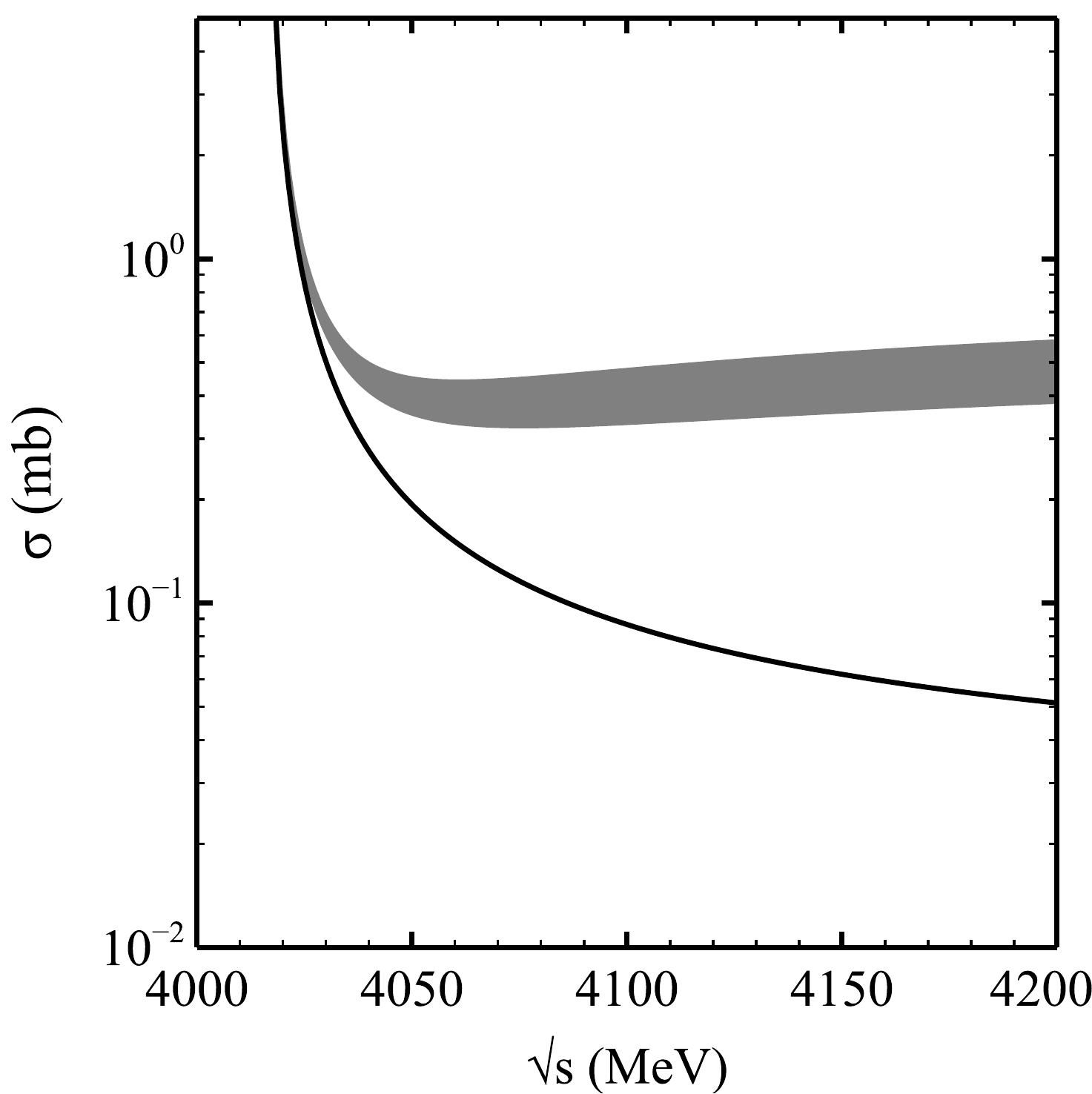}
\caption{(Left) Cross section for the reaction $\bar D^* D\to\pi X$. The solid line represents the result for the $\bar D D\to \pi X$ cross section, and we have shown it for the purpose of comparison. The dashed line represents the result for the cross section of the process $\bar D^* D\to\pi X$ considering only the $t$ channel diagram in Fig.~\ref{DbarstarD}. The shaded region is the result obtained with both $t$ and $u$ channel diagrams of Fig.~\ref{DbarstarD} considering cut-offs in the range 700-1000 MeV. (Right) Cross section for the reaction $\bar D^* D^*\to\pi X$. The solid line represents the cross section without the contribution from the diagrams in Fig.~\ref{DstarDstar}b and \ref{DstarDstar}d, which contain the vertex $X\bar D^* D^*$. The shaded region represents the result for the cross section when including the contribution of all the diagrams in Fig.~\ref{DstarDstar}. }\label{crossDbarstarD}
\end{figure}

 In Fig.~\ref{crossDbarstarD} (left side) we show the cross section related to the process $\bar D^* D\to \pi X$. The solid line represents the result for the $\bar D D\to \pi X$ cross section, which does not involve any anomalous vertex, and it is shown for comparison. The dashed line is the cross section for the $\bar D^* D\to \pi X$ process without considering the diagrams involving the anomalous vertex $X\bar D^* D^*$, i.e., only with the $t$ channel diagram shown in Fig.~\ref{DbarstarD}c. The shaded region represents the result found with both $t$ and $u$ channel diagrams shown in Figs.~\ref{DbarstarD}c and ~\ref{blob} (with the latter ones involving the $X\bar D^* D^*$ vertex) when changing the cut-off needed to regularize the loop integrals in the range 700-1000 MeV. As can be seen, the results do not get very affected by a reasonable change in the cut-off. Clearly, the vertex $X\bar D^* D^*$ plays an important role in the determination of the $\bar D^* D\to \pi X$ cross section, raising it by around a factor 100-150. We also show that the cross section for the reaction $\bar D^* D\to\pi X$ is larger than that for $\bar D D\to\pi X$ and, thus, the consideration of the former reaction in a calculation of the abundance of the $X$ meson in heavy ion collisions could be important.

In Fig.~\ref{crossDbarstarD} (right side) we show the results for the cross section of the reaction $\bar D^* D^*\to \pi X$. The solid line corresponds to the result found without the anomalous $X\bar D^* D^*$ contribution, while the shaded region is the result considering the diagrams involving this anomalous vertex. This region has been obtained by using an effective Lagrangian to describe the $XD^*\bar D^*$ vertex. To estimate the coupling $g_{X \bar D^* D^*}$ present in this effective Lagrangian we have used the results obtained in Fig.~\ref{crossDbarstarD} (right side), which lead to $g_{X\bar D^* D^*}\sim 1.95\pm 0.22$. We refer the reader to Ref.~\cite{our} for more details on this calculation.

As can be seen, the contribution from the diagrams involving the $X\bar D^* D^*$ vertex is important, raising the cross section about a factor 8-10. 

Therefore, as in case of the $\bar D^* D\to\pi X$ reaction, the consideration of the anomalous vertices could play an important role when determining the $X$ abundance in heavy ion collisions.

\section{Summary}\label{Sum}

In this work we have obtained the production cross sections of the reactions $\bar D D$, $\bar D^* D$, $\bar D^* D^*\to\pi X$, considering $X(3872)$ as a molecular state of $\bar D D^*-\textrm{c.c}$. 
We have shown that the inclusion of the anomalous vertex $X\bar D^* D^*$ is very important  for the processes $\bar D^* D\to\pi X$ and $\bar D^* D^*\to\pi X$ and it could play an important role in the determination of the abundance of 
the $X$ meson in heavy ion collisions.

\section{Acknowledgements}
The authors would like to thank the Brazilian funding agencies FAPESP and CNPq for the financial support. \\

\end{document}